\documentclass{aa}  
\usepackage{graphicx}
\usepackage[varg]{txfonts}
\usepackage{color}

\begin{document} 
   \title{Ambipolar diffusion regulated collapse of filaments threaded by 
	perpendicular magnetic fields}

   \author{C. A. Burge \inst{1} \and S. Van Loo \inst{1}
	\and S. A. E. G. Falle\inst{2} \and T. W. Hartquist \inst{1}}
   \authorrunning{Burge et al.}

   \offprints{C. A. Burge\\ \email{c.a.burge@leeds.ac.uk}}

   \institute{School of Physics and Astronomy, University of Leeds, Leeds LS2 9JT, UK
	\and Department of Applied Mathematics, University of Leeds, Leeds LS2 9JT, UK}
          
   \date{Received date; accepted date}

   \abstract
	{In giant molecular clouds (GMCs), the fractional ionisation is low enough that the 
	neutral and charged particles are weakly coupled. A consequence of this is that the 
	magnetic flux redistributes within the cloud, allowing an initially magnetically 
	supported region to collapse.}
	{We aim to elucidate the effects of ambipolar diffusion on the evolution of 
	infinitely long filaments and the effect of decaying turbulence on that 
    	evolution.}
	{First, in ideal magnetohydrodynamics (MHD), a two-dimensional cylinder of an 
	isothermal magnetised 
	plasma with initially uniform density was allowed to evolve to an equilibrium 
	state. Then, the response of the filament to ambipolar diffusion was followed 
	using an adaptive mesh refinement multifluid MHD code. Various ambipolar 
	resistivities were chosen to reflect different ratios of Jeans length to 
	ambipolar diffusion length scale. To study the effect of turbulence on 
	the ambipolar diffusion rate, we perturbed the equilibrium filament with 
	a turbulent velocity field quantified by a rms sonic Mach number, 
	$M_{\rm rms}$, of 10, 3 or 1.}
	{We numerically reproduce the density profiles for filaments 
	that are in magnetohydrostatic and pressure equilibrium with their 
	surroundings obtained in a published model and show that these equilibria 
	are dynamically stable. If the effect of ambipolar diffusion is considered,
        these filaments lose magnetic support initiating cloud collapse. The 
	filaments do not lose magnetic flux. Rather the magnetic flux is redistributed
        within the filament from the dense centre towards the diffuse envelope.

        The rate of the collapse is inversely proportional to the fractional 
	ionisation and two 
        gravitationally-driven ambipolar diffusion 
	regimes for the collapse are observed as predicted in a published model. 
	For high values of the ionisation coefficient, 
	that is $X \geq 10^{-7}$, the gas is strongly coupled to the magnetic field
        and the Jeans length is larger than the ambipolar diffusion length scale.
        Then the collapse is 
        governed by magnetically-regulated ambipolar diffusion.
	The gas collapses at velocities much lower than the sound 
	speed. For $X \lesssim 10^{-8}$, the gas is weakly coupled to the magnetic
	field and the magnetic support is removed by 
	gravitationally-dominated
  	ambipolar diffusion. Here, neutrals and ions only collide sporadically,
	that is the ambipolar diffusion length scale is larger than the Jeans length,
	and the gas can attain high collapse velocities.  
        
        When decaying turbulence is included, additional support is provided 
        to the filament. This slows down the collapse of the filament even in the 
	absence of a magnetic field. When a magnetic field is present, the collapse 
	rate increases by a ratio smaller than for the non-magnetic case. This is 
	because of a speed-up of the 
	ambipolar diffusion due to larger magnetic field gradients generated by 
	the turbulence and because the ambipolar diffusion aids the dissipation of
	turbulence below the ambipolar diffusion length scale. The highest increase
	in the rate is observed for the lowest ionisation coefficient and the highest
	turbulent intensity.}

   \keywords{Methods: numerical -- ISM: magnetic fields -- ISM: clouds -- 
	Stars: formation -- MHD}

   \maketitle
   
\section{Introduction}
Giant molecular clouds (GMCs) contain regions of enhanced density where star 
formation occurs. These regions often take the form of structures such as clumps 
and filaments \citep[e.g.][]{1994ApJ...425..641S,2003ApJS..149..343E}. These 
structures may be thermally supported, or in cases where the mass of the object 
exceeds the Jeans mass, a magnetic field can provide support against gravitational 
collapse, provided that it is sufficiently strong.  Only if such structures are able 
to fragment or collapse into sufficiently dense cores, can protostellar objects form.
Various mechanisms for initiating this collapse have been suggested, including 
collisions between clouds \citep[e.g.][]{2014ApJ...792...63T}, shock-cloud interactions 
\citep[e.g.][]{2006MNRAS.365...37B, 2013MNRAS.433.1258V}, and perturbation of cores 
by waves \citep[e.g.][]{1995ApJ...440..686M, 2007MNRAS.376..779V}.

\citet{1956MNRAS.116..503M} suggested that ambipolar diffusion due to the relative motion 
between the neutrals and ions, can also fragment molecular clouds into dense cores. 
Detailed calculations of \citet{1976ApJ...207..141M, 1979ApJ...228..475M} subsequently showed that 
ambipolar diffusion does, indeed, leads to the self-initiated collapse of dense central regions while 
the cloud envelope remains magnetically supported as magnetic flux is redistributed. 
Additional numerical simulations of magnetically sub-critical self-gravitating sheets or layers 
further confirm the ambipolar-diffusion regulated fragmentation process 
\citep[e.g.][]{2011ApJ...728..123K,2014ApJ...794..127K}.

In this paper we consider the effect of ambipolar diffusion and velocity perturbations 
on magnetically sub-critical filaments. In these models the magnetic field is 
perpendicular to the filament axis. 
Filamentary clouds threaded by magnetic fields (both parallel and perpendicular to their 
axis) are expected to form due to gravitational and thermal instabilities within thin 
dense layers \citep[e.g.][]{2007MNRAS.380..499K,2011MNRAS.414.2511V,2014ApJ...789...37V}. 
Many observed filaments in GMCs exhibit such a configuration \citep{2013MNRAS.436.3707L}, 
but as yet little theoretical study of such structures has been performed. 
In Sect.~\ref{Model} we describe the numerical code 
and our initial conditions based on the analytic work of \citet{2014ApJ...785...24T}. 
Then, in Sect.~\ref{ambipolar}, we investigate the effect of ambipolar diffusion on 
the evolution of the filaments. We also examine the interaction of velocity 
perturbations with the diffusion process in Sect.~\ref{sect:turbulent}. Finally, in 
Sect.~\ref{summary}, we discuss and summarise our results.

\section{The model}\label{Model}
\subsection{Multifluid code}
Within molecular clouds the fractional ionisation is low, so the gas can be treated 
as a multispecies fluid consisting of neutrals, electrons and ions. Furthermore, 
we have adopted an isothermal equation of state $p = \rho T$ for all fluids with $T$ the 
isothermal temperature.  For the neutral fluid, the governing isothermal equations 
are given by 
\begin{align}
    \frac{\partial{\rho_n}}{\partial{t}} + \nabla (\rho_n {\bf v}_n)  =  0,\\
    \frac{\partial{\rho_n {\bf v}_n}}{\partial{t}} + \nabla  (\rho_n {\bf v}_n
        {\bf v}_n + p_n)  =  {\bf J}\times{\bf B} - \rho_n \nabla \phi, 
\end{align}
with $\rho_n$ the neutral density, ${\bf v}_n$ the neutral velocity, $p_n$ the neutral
pressure, $\phi$ the gravitational potential, $\bf{B}$ the magnetic field and $\bf{J}$ 
the current given by $\bf{J} = \nabla \times \bf{B}$. As we are interested in 
the filament configuration and evolution for a given line mass and mass-to-flux ratio, 
we have only applied the gravitational force $\rho_n \nabla \phi$ for filament gas to avoid 
accretion of external gas by the filament. In the limit of small mass densities for 
the electrons and ions, the gravitational potential can be calculated using the 
Poisson equation  
\begin{equation}
\nabla^2 \phi=4\pi G\rho_n.
\end{equation}
Here $G$ is the gravitational constant which we set to 1.

For the charged fluids, we assumed ionisation equilibrium \citep[e.g.][]{1979ApJ...231..372E} 
and also neglected their inertia so that the equations reduce to
\begin{eqnarray}\label{eq:charged}
     \rho_{j} = 30 X \sqrt{\rho_{n}},\\
     \alpha_{j} \rho_{j} ({\bf E} + {\bf v}_j\times {\bf B}) + 
        \rho_{j} \rho_n K_{jn} ({\bf v}_n - {\bf v}_j) =  0. \nonumber
\end{eqnarray}
where $X$ is the ionisation coefficient related to the ionisation fraction $\chi = X \rho_n^{-1/2}$,
$\alpha_j$ the charge-to-mass ratio, $K_{jn}$ the 
collision coefficient of the charged fluid $j$ with the neutrals, and $\bf{E}$ 
the electric field. Here $j$ stands for either the electrons or ions.  
We used 
$10^{-6} \le X \le 10^{-8}$ and adopted a charge-to-mass ratio 
$\alpha_e = -8.39\times 10^{15}$ for the electrons and 
$\alpha_i = 1.52\times 10^{11}$ for the ions, and $K_{en} =2.99 \times 10^{10}$ 
and $K_{in} = 2.06 \times 10^{7}$ as the collision coefficients between the electrons 
and ions with the neutrals. We assumed an ion mass of $30m_H$. 

If we take a dimensional sound speed within the cloud of 0.35 km s$^{-1}$ 
(corresponding to $\approx 20$K), this equates to a cloud radius of 
$\approx 0.3$pc \citep[consistent with observed filament widths;][]{2010A&A...518L.102A}. 

We have also included the effects of ambipolar diffusion. Using the charged fluid 
momentum equation to substitute ${\bf E}$ in the Maxwell-Faraday equation, 
the evolution of the magnetic field is governed by 
\begin{equation}\label{eq:Bfield}
         \frac{\partial{{\bf B}}}{\partial{t}} - \nabla \times       
        ({\bf v}_n \times {\bf B}) = \nabla \times \left(r_a 
        \frac{((\nabla \times {\bf B}) \times {\bf B})\times {\bf B}}{B^2} \right),
\end{equation}
where we have neglected the contributions of the resistivity along the field and the 
Hall resistivity. This can be done as the Hall parameter for electrons and ions are much 
larger than unity.
Furthermore, as the Hall parameter for electrons is larger than 
that for ions, $r_a$ (the ambipolar resistivity) is given by \citep{2003MNRAS.344.1210F}
\begin{equation}\label{eq:ra}
        r_a = \frac{B^2}{\rho_i \rho_n K_{in}} = \frac{B^2}{30 X \rho^{3/2}_n K_{in}}.
\end{equation}

These equations are solved using the multifluid version of the adaptive mesh 
refinement code MG, which is described in detail in \citet{2008A&A...484..275V} and 
based on the algorithms outlined in \cite{2003MNRAS.344.1210F}. This scheme uses a 
second-order Godunov solver with a linear Riemann solver for the neutral fluid 
equations. The charged fluid velocities can be calculated from the reduced momentum 
equation and the magnetic field is advanced explicitly which imposes an extra 
restriction on the stable time step, besides the Courant condition, at high numerical 
resolution due to the ambipolar resistivity term, that is $\Delta t < \Delta x^2/4r_a$ 
\citep{2003MNRAS.344.1210F}. The Poisson equation for the self-gravity is solved 
using a full approximation multigrid.

The code uses a hierarchical grid structure in which the grid spacing of level 
$n$ is $\Delta x/2^n$, where $\Delta x$ is the grid spacing of the coarsest level. 
The coarsest grids cover the entire domain, but higher level grids do not necessarily. 
A divergence cleaning algorithm is used to eliminate errors arising from 
non-zero $\nabla \cdot \mathbf{B}$ \citep{2002JCoPh.175..645D}.

  \begin{table}
  \caption{Initial conditions given in dimensionless units along with the 
	resolution of each model and the name of the corresponding Tomisaka model.}
  \label{table:values}      
  \centering                         
  \begin{tabular}{c c c c c c }        
  \hline\hline                 
  Tomisaka & $\rho_0$ &$B_0$& $p_{\rm ext}$ &  $\rho_{\rm ext}$   & Resolution \\
  Model    &          &     &               &                     & (in x $\&$ y)\\
  \hline                       
     Aa  & 0.327 & 1.152 &  1.99 $\times 10^{-2}$ & 0.001 &  1024    \\   
     Ab  & 0.631 & 1.152 & 1.99 $\times 10^{-2}$ & 0.001 &  1024    \\
     Ac  & 0.800 & 1.152 &  1.99 $\times 10^{-2}$ & 0.001 &  2048  \\
     C3a & 0.547 & 0.798 & 0.318 & 0.01 &  1024  \\
     C3b & 0.699 & 0.798 & 0.318 & 0.01 &  1024  \\ 
     C3c & 0.719 & 0.798 & 0.318 & 0.01 &  2048 \\     
     D1  & 1.033 & 1.995 & 1.989 & 0.1 &  4096 \\
     D2  & 2.176 & 6.308 & 1.989 & 0.1   & 4096 \\
     D3  & 4.281 & 19.95 & 1.989 & 0.1   & 4096  \\
  \hline                                   
  \end{tabular}
  \end{table}

\subsection{Initial conditions}
For our initial conditions we have assumed infinitely long, isothermal, magnetised 
filaments that were initially in magnetohydrostatic equilibrium and in pressure 
equilibrium with the external medium. \citet{2014ApJ...785...24T} analytically derived 
the density profiles and magnetic field structures of such filaments and showed 
that the magnetohydrostatic configurations depend strongly on the centre-to-surface 
density contrast, the ratio of the magnetic-to-thermal pressure of the external 
medium, and the radius of the cloud. While we can easily adopt his method to produce 
our initial conditions, we chose to reproduce the different filament configurations 
numerically.  Therefore, we considered a cylinder along the $z$-axis with a uniform 
density $\rho_0$ and radius $R_0$ and threaded by a uniform magnetic field $B_0$ in the 
$y$-direction. Our model parameters, $\rho_0$ and $B_0$, are listed in 
Table~\ref{table:values}, along with the Tomisaka model they represent, while we 
assumed the filament radius $R_{0}$ and temperature $T_0$ both to be 1. (Our 
model parameters are dimensionless.) Instead of varying the initial cloud radius, 
we varied the external pressure to produce results consistent with Tomisaka's work. 
Each model is thus defined by its line mass, mass-to-flux ratio, and 
external pressure.

The filament is embedded in a diffuse medium. This external medium has a density 
$\rho_{\rm ext}$ much lower than the filaments so that the gravitational potential
is determined solely by the filament. Furthermore, as we assumed pressure-equilibrium, 
the external temperature is given by $T_{\rm ext} = \rho_0/\rho_{\rm ext}$.  We used 
a computational domain $-5 \le x \le 5$, $-5 \le y \le 5$ with the finest grid spacing 
smaller than the Jeans length to avoid artificial fragmentation 
\citep{1997ApJ...489L.179T}. The highest resolution for each model is given in 
Table \ref{table:values}.

  \begin{figure}
  \centering
  \includegraphics[width=9cm]{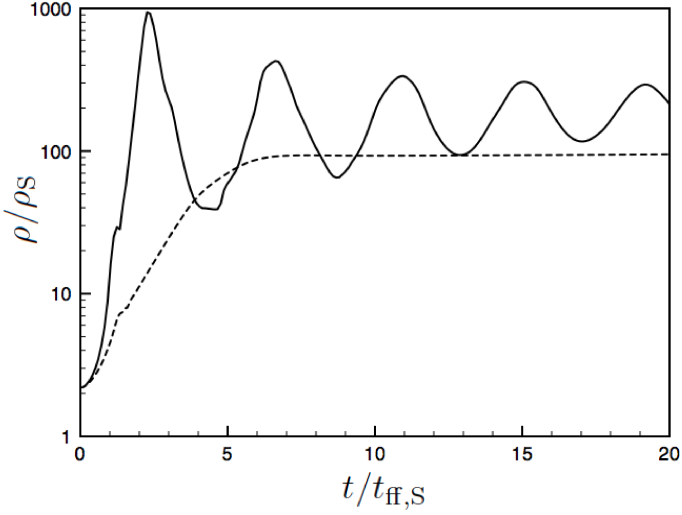}
  \caption{Normalised maximum density, $\rho/\rho_{\rm S}$ as a function of 
  the time for model C3b. The solid line is for the undamped evolution, while
  the dashed line includes drag (with $C=5$).}
  \label{fig:evolution}
  \end{figure}

  \begin{figure*}
  \centering
  \includegraphics[width=\textwidth]{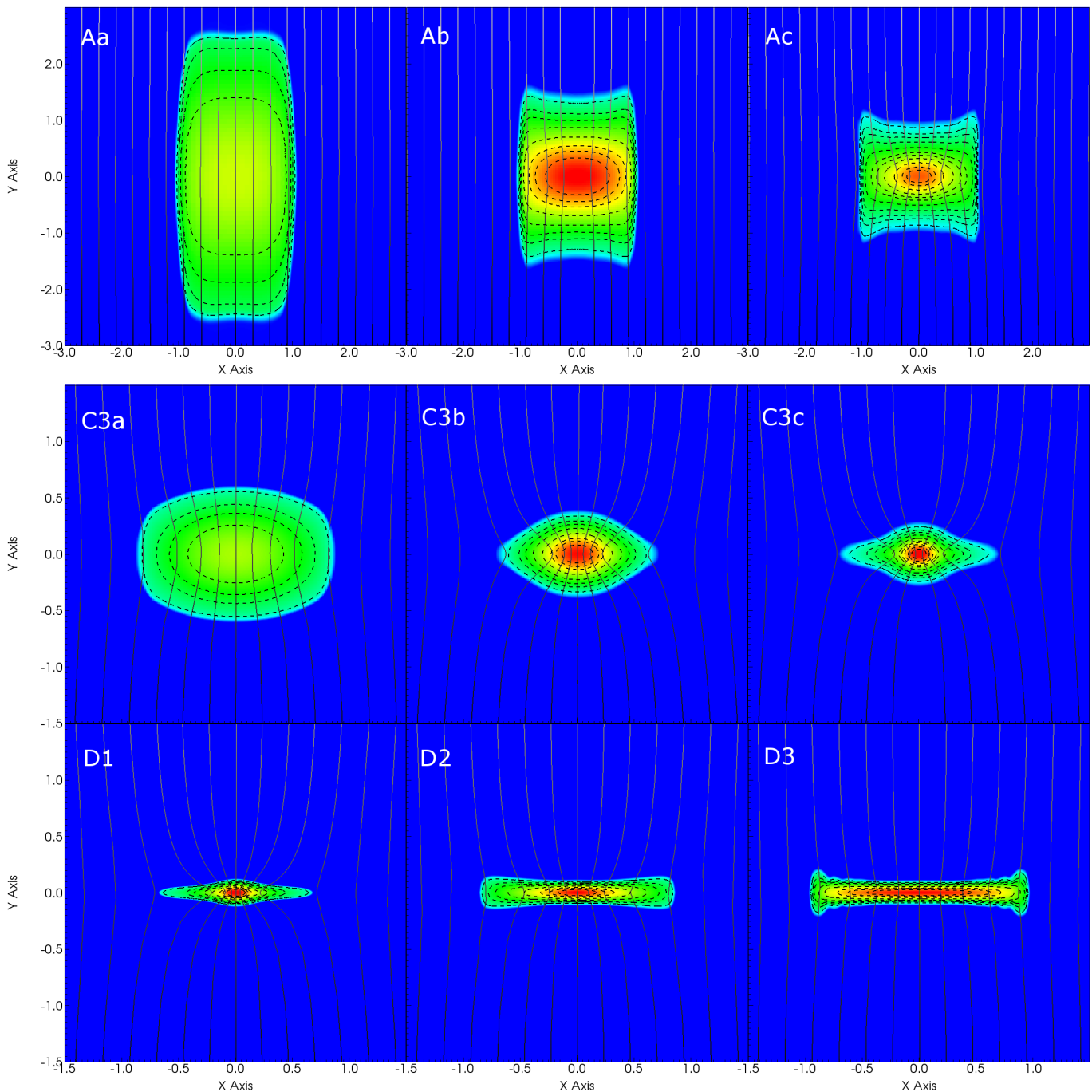}
  \caption{Normalised logarithmic density, $\rho/\rho_{\rm S}$, of the 
  equilibrium configuration for the models listed in Table~\ref{table:values}. 
  The density range is from 0.1 to 100, except for model Ac and C3c which 
  have a maximum density of 1000 and 300, respectively. The dashed lines show 
  contour lines for density values of 1, 2, 3, 5, 10, 20, 30, 50, 100, 200, 300 and 500. 
  The magnetic field lines are shown by the solid lines.}
  \label{fig:equilibrium}
  \end{figure*}

  \begin{table}
   \begin{center}
  \caption{Magnetic flux per unit length given in dimensionless units along with the 
	ratio of the line mass to the critical values given in 
	Eqs.~\ref{eq:tomisaka}~and~\ref{eq:ostriker} for each model. For model D1 
	and D2, the ratio exceeds unity marginally, but we should remember that Eq.~\ref{eq:tomisaka}
        is only an empirically-derived limit and some variation is expected.}
  \label{table:flux_ratios}      
  \centering                         
  \begin{tabular}{c c c c c c }        
  \hline\hline                 
  Tomisaka & $\phi_{cl}$ & $\lambda/\lambda_{\rm th, max}$ & $\lambda/\lambda_{\rm max}$ \\
  Model    &          &     &               \\
  \hline                       
     Aa  & 2.304 & 0.514 &  0.382 \\
     Ab  & 2.304 & 0.992 &  0.737 \\
     Ac  & 2.304 & 1.257 &  0.934 \\
     C3a & 1.596 & 0.859 & 0.693 \\
     C3b & 1.596 & 1.098 & 0.886 \\
     C3c & 1.596 & 1.129 & 0.911 \\
     D1 & 3.990 & 1.623 & 1.015 \\
     D2 & 12.62 & 3.428 & 1.180 \\
     D3 & 39.90 & 6.725 & 0.963 \\
  \hline                                   
  \end{tabular}
  \end{center}
  \end{table}

\section{Equilibrium filaments}\label{tomisaka}
For the model parameters listed in Table~\ref{table:values} the resulting 
filaments are all magnetically sub-critical as their line mass, $\lambda$,
is  below the maximum value \citep{2014ApJ...785...24T}
\begin{equation}\label{eq:tomisaka}
\lambda_{\rm max} \approx 0.30 \frac{\phi_{cl}}{G^{1/2}}+ \frac{2c^2}{G}
\end{equation}
where $\phi_{cl}$ is the magnetic flux per unit length and $c = \sqrt{T}$
the thermal sound speed. The magnetic flux is here $\phi_{cl} = 2 R_0B_0 = 2B_0$.
We note that Eq.~\ref{eq:tomisaka} differs from the Tomisaka expression due to 
incorporation of a factor of $\sqrt{4\pi}$ in the magnetic field and the 
definition of the magnetic flux per unit length. Also, the thermal pressure
contribution is set to the critical value derived by \citet{1964ApJ...140.1056O}
\begin{equation}\label{eq:ostriker}
\lambda_{\rm th,max} = \frac{2c^2}{G}.
\end{equation}
For all of our models the magnetic field thus provides enough support 
to avoid gravitational collapse, but some of the models, that is Aa, Ab and C3a, 
are also thermally sub-critical as their line mass is below the critical 
value $\lambda_{\rm th,max}$ 
(see Table~\ref{table:flux_ratios}).

Sub-critical filaments evolve towards their magnetohydrostatic equilibrium if no 
other forces are considered.  We follow this evolution using the multifluid MHD code
with the ambipolar resistivity set to zero, so that Eq.~\ref{eq:Bfield} becomes 
\begin{equation}
    \frac{\partial{{\bf B}}}{\partial{t}} - \nabla \times       
        ({\bf v}_n \times {\bf B}) = 0,
\end{equation}
Then the neutral fluid equations together with the magnetic field equation reduce to 
the ideal MHD equations. However, the momentum equation (see Eq.~2) 
includes the Lorentz force as a source term and is thus not in its conserved form, that is 
\begin{equation}
\frac{\partial{\rho_n {\bf v}_n}}{\partial{t}} + \nabla  (\rho_n {\bf v}_n
        {\bf v}_n + p_n + \frac{B^2}{2} - {\bf B} {\bf B})  =  - \rho_n \nabla \phi.
\end{equation}
Because the momentum equation can be written in different forms, it is possible
to formulate two distinct numerical approaches. An ideal MHD code uses the conserved 
momentum equation and one can construct a Riemann problem combining all flow variables
\citep[e.g.][]{1988JCoPh..75..400B}. The multifluid MHD code only solves a 
non-magnetic Riemann problem for the advection of the density and velocity while the 
magnetic field is advected separately \citep[][]{2003MNRAS.344.1210F}. This latter 
approach is simpler, but also less accurate than the former one when applied to ideal MHD. 
Nevertheless, the equilibrium filamentary structures
calculated with both the ideal and multifluid version of MG are qualitatively 
identical with central densities differing only by a few percent.

The initially uniform filaments undergo gravitational contraction as there is no 
thermal or magnetic pressure gradient to counter self-gravity. This pressure gradient 
is established quickly, but the inertia of the gas causes the filaments to oscillate 
around their magnetohydrostatic equilibrium (see for example Fig.~\ref{fig:evolution}). 
To hasten the evolution towards the equilibrium, we damp the neutral velocity by 
introducing a drag force, which we decrease to zero as the simulation approaches 
equilibrium. 
This is done by adding a drag force term, that is $-C \rho_n \bf{v}_n$, on the left 
hand side of neutral momentum equation with $C$ the drag coefficient. 
Figure~\ref{fig:evolution} shows the temporal evolution of the central density with 
time for model C3b. The central density increases by nearly two orders of magnitude
to its equilibrium value, that is from $\rho = 0.699$ to  $100 \rho_{\rm S} \approx 32$  
with $\rho_{\rm S} = p_{\rm ext}/T_0$ the surface density of the filament within 
$\approx 5 t_{\mathrm {ff,S}}$ (where 
$t_{\mathrm {ff,S}} = \left(1/4 \pi G\rho_{\rm S}\right)^{1/2}$ is the free-fall time
evaluated with the surface density). Contrary to the undamped evolution, the filament does not 
oscillate as it attains it equilibrium configuration. Once the central density remains 
constant for several free-fall time we say the filament has reached its equilibrium 
configuration. The undamped solution has a higher central density 
that continues to increase. This is due to numerical diffusion during the contraction
phase which causes the cloud to lose magnetic flux and leads to a denser filament 
centre than expected from its initial conditions.

Figure~\ref{fig:equilibrium} shows the equilibrium density and magnetic field 
configuration for the models listed in Table~\ref{table:values}. Comparing these 
structures with the analytic results of \cite{2014ApJ...785...24T}, we find 
identical density and magnetic field structures and peak central densities.
Models A and C3 show the effect of increasing filament mass, while models D show
the effect of an increasing magnetic field strength. We also examine whether these 
equilibrium structures are stable. Therefore, we superimpose 
a turbulent velocity field with a rms sonic Mach number of 0.3 (see 
Sect.~\ref{sect:turbulent} for implementation details). We do not consider 
driven turbulence for this stability study.  While the density structure is modified 
initially by the velocity perturbations, the filaments attain their original 
equilibrium configuration as the turbulence decays. Our results then not only confirm 
Tomisaka's analytic results; they also show that the equilibria are stable. 

  \begin{figure}
  \centering
  \includegraphics[width=9cm]{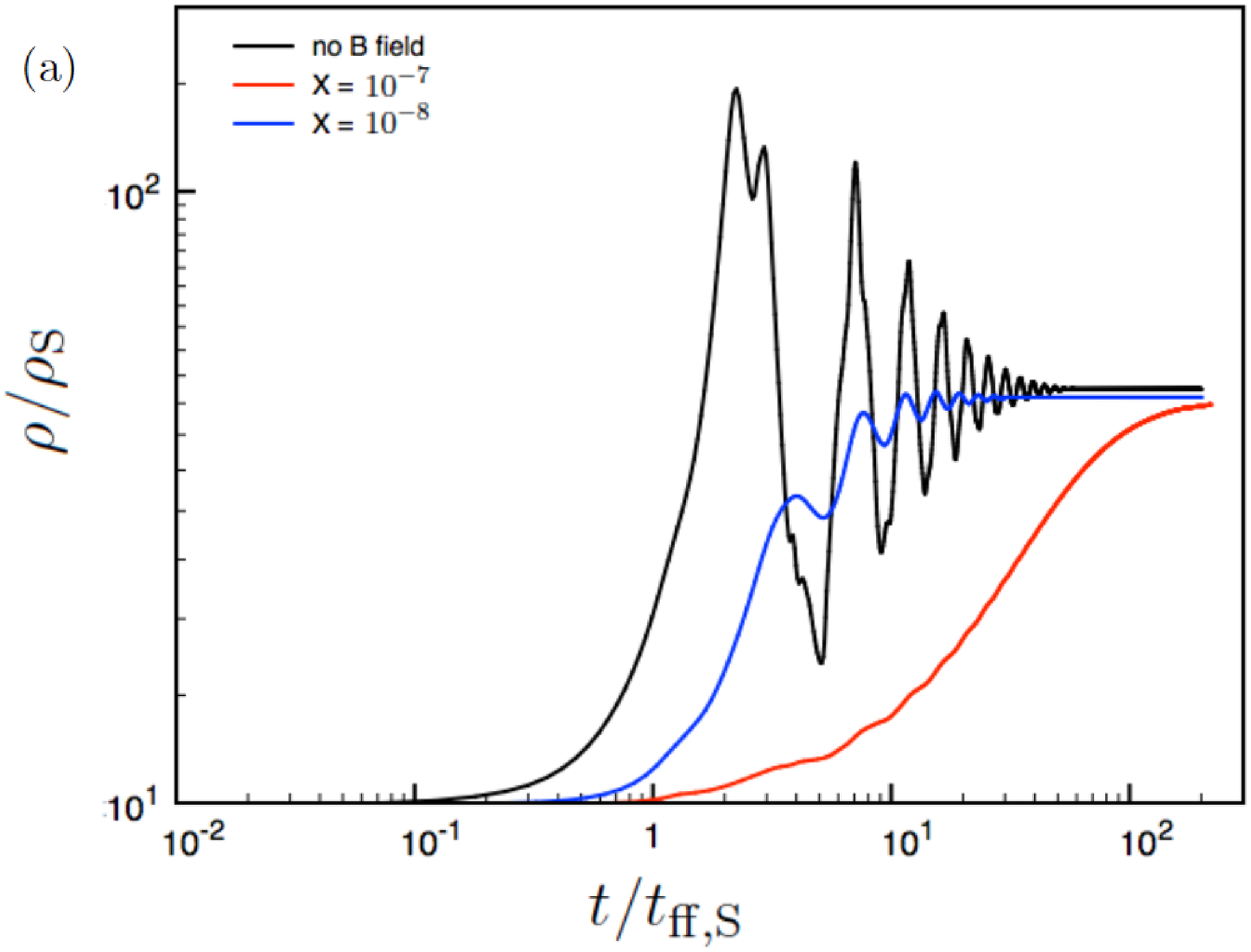}
  \includegraphics[width=8.5cm]{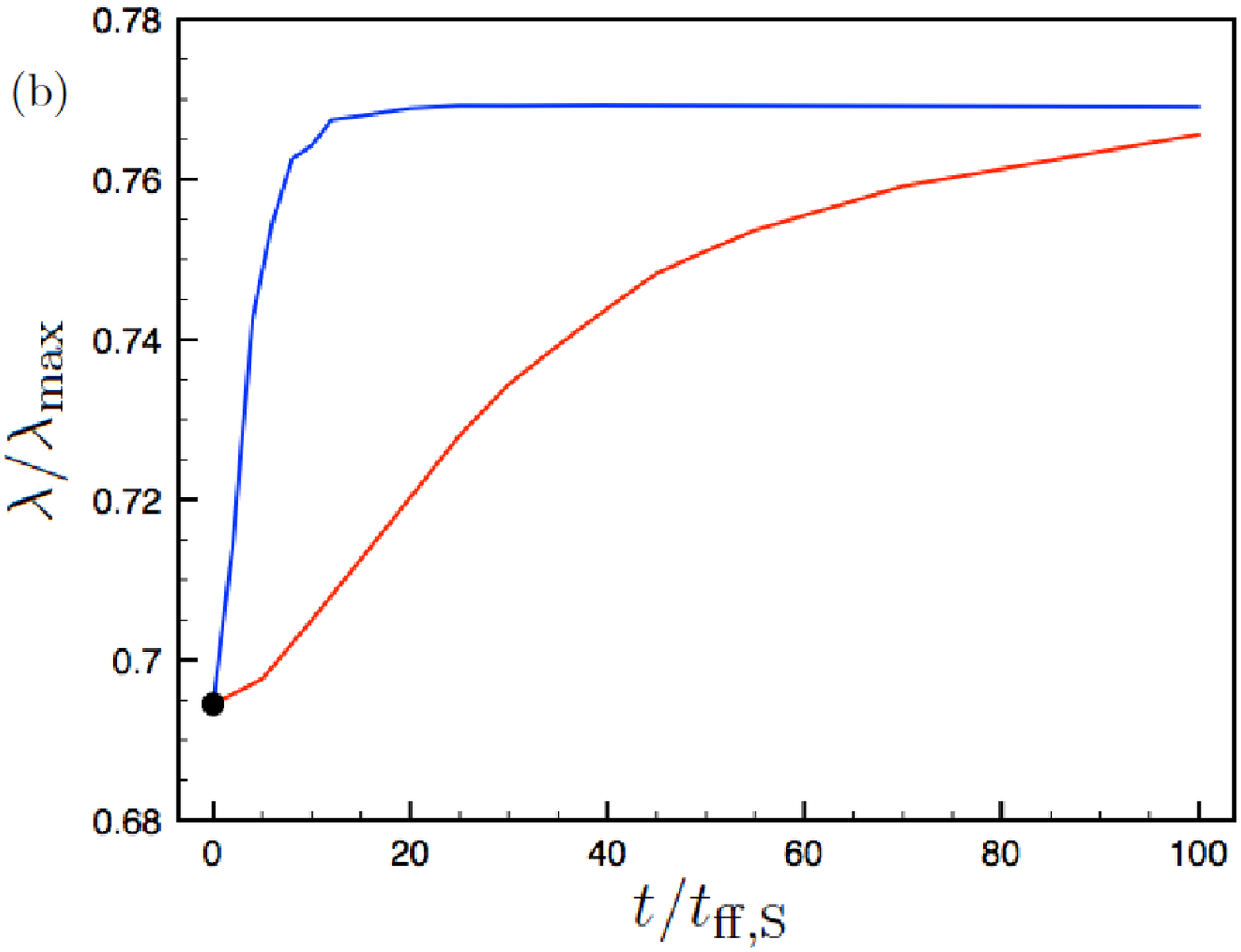}
  \caption{Ambipolar diffusion regulated collapse for model C3a. 
  The blue and red line show the evolution for $X= 10^{-7}$ and $X = 10^{-8}$
  respectively, while the black line shows the evolution without magnetic fields. 
  Panel (a) shows the maximum (i.e. central) density, while panel (b) shows the 
  ratio of the line-mass to the critical value given by Eq.~\ref{eq:tomisaka}. The
  black dot shows the value for the equilibrium structure.
  }
  \label{fig:C3a evolution}
  \end{figure}
 
\section{Ambipolar diffusion}\label{ambipolar}
The magnetohydrostatic filament configurations derived in the previous
section are magnetically supported against gravitational collapse. 
Gravitational collapse of the filament is then possible only if the line mass 
increases above the critical value given by Eq.~\ref{eq:tomisaka}. This can
only happen if the magnetic flux decreases. In a weakly ionised plasma, ambipolar 
diffusion, or ion-neutral drift causes structural reorganisation of the magnetic 
field. When ambipolar diffusion occurs, the magnetic field diffuses out of the 
filament, thereby decreasing the magnetic flux. An initially magnetically sub-critical 
filament then becomes super-critical.

  \begin{figure}
  \includegraphics[width=9cm]{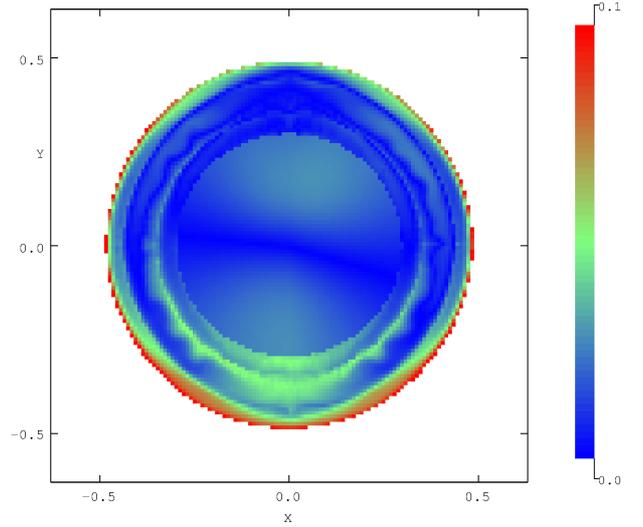}
  \caption{Fractional difference between model C3a with $X = 10^{-8}$ and the 
	Ostriker hydrostatic equilibrium profile with the same central density.}
  \label{fig:deviation Ostriker}
  \end{figure}

\subsection{Thermally sub-critical filaments}
As mentioned earlier, some of the filament models, that is Aa, Ab and C3a, are thermally 
sub-critical. As these filaments lose magnetic flux, their density structure is
expected to evolve towards a new equilibrium state supported solely by thermal 
pressure gradients.  
 
Figure~\ref{fig:C3a evolution} shows the effect of ambipolar diffusion on model C3a.
Due to the gradients of the magnetic field, ambipolar diffsuion is initiated and the filament loses 
magnetic support against self-gravity
(see Panel (b)) and, consequently, the filament starts to contract (see Panel (a)). 
As the ambipolar diffusion coefficient is inversely proportional to the fractional ionisation, 
the contraction rate increases with decreasing fractional ionisation. However, the contraction 
does not continue indefinitely. The central density only increases by a factor of 
approximately five, after which it remains constant and, hence, the filament reaches 
a new stable 
configuration. 
The magnetic flux per unit length also reaches a new stable value indicating that 
the magnetic field is now uniform and can no longer provide support against self-gravity.
Although the filament is now magnetically super-critical, it is still thermally 
sub-critical.  Then the filament's density profile is given by the hydrostatic 
equilibrium profile
\citep{1964ApJ...140.1056O}
\begin{equation}\label{ost}
\rho(r)=\frac{\rho_0}{[1+(r/R_0)^2]^2}, 
\end{equation}
where $R_0 = \sqrt{2c^2/\pi G\rho_0}$ is the thermal scale height. 
By integrating Eq.~\ref{ost} to obtain the line mass $\lambda$ and taking
$\lambda = 1.718$ and $p_{\rm ext} = 0.318$ (see parameters for model C3a),
one can show that the central density of the equilibrium cylinder 
needs to be $\rho_0 = 16.06 \approx 50\rho_{\rm S}$. All ambipolar-diffusion regulated 
models attain values for $\rho_0$ that are nearly equal to this value
(see Fig.~\ref{fig:C3a evolution}). 
As a test we also run a model with the magnetic field removed from the equilibrium 
configuration. This model also reproduces a similar central density, so that 
ambipolar diffusion clearly removes the magnetic support from the filament.

Not only do we have the same central density in the simulation, model C3a 
can be entirely fitted with the analytic profile given by Eq.~\ref{ost}. 
Figure~\ref{fig:deviation Ostriker} shows the relative difference between 
the equilibrium density profile attained with an ambipolar diffusion model 
using $X = 10^{-8}$ and the Ostriker density distribution with the same central
density. The relative difference for most of the cylindrical filament is 
below 5\%; the error approaches 10\% only near the edge.

Similar results are obtained for the other thermally sub-critical models Aa 
and Ab.

  \begin{figure}
  \centering
  \includegraphics[width=9cm]{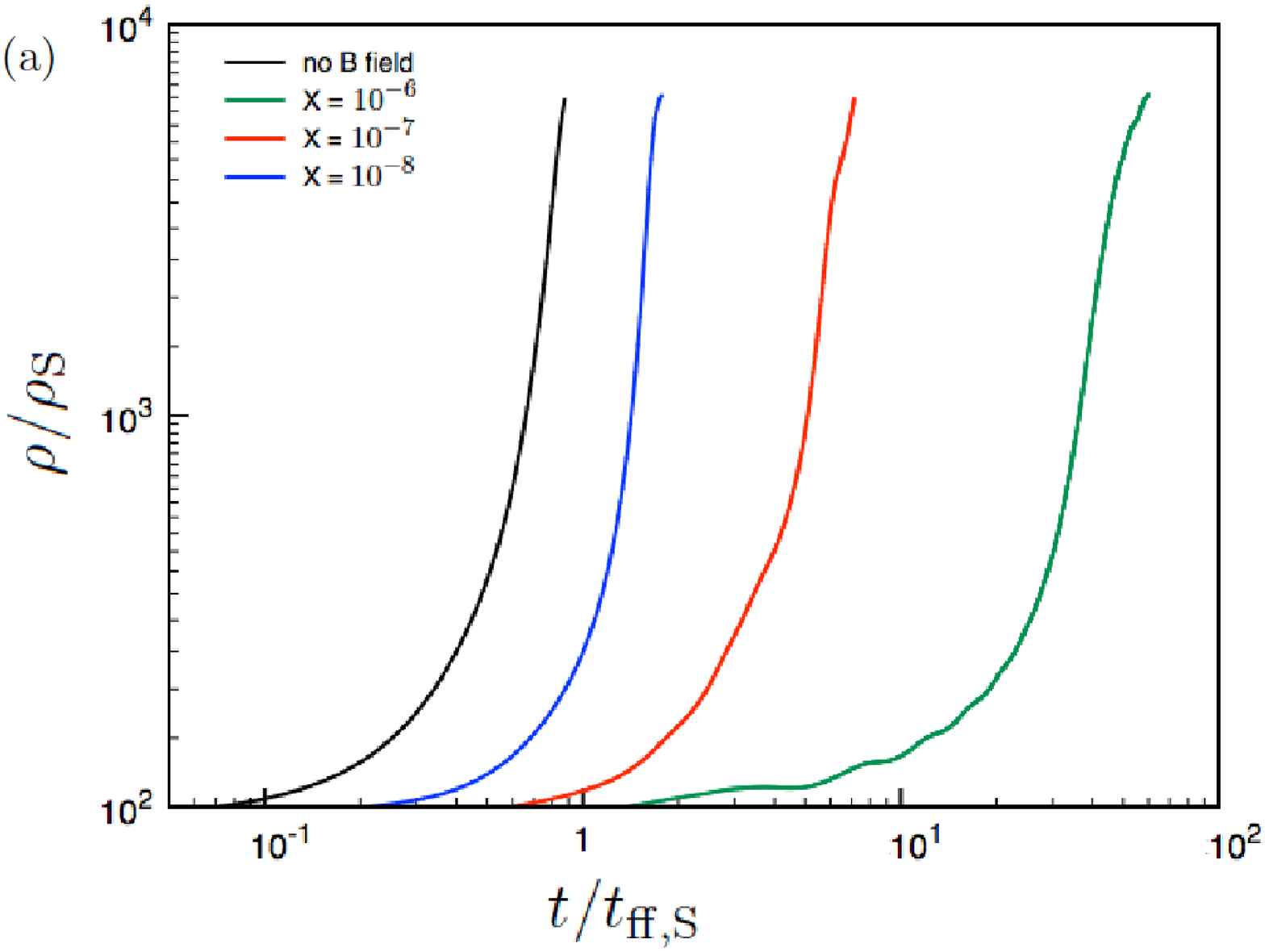}
  \includegraphics[width=8.5cm]{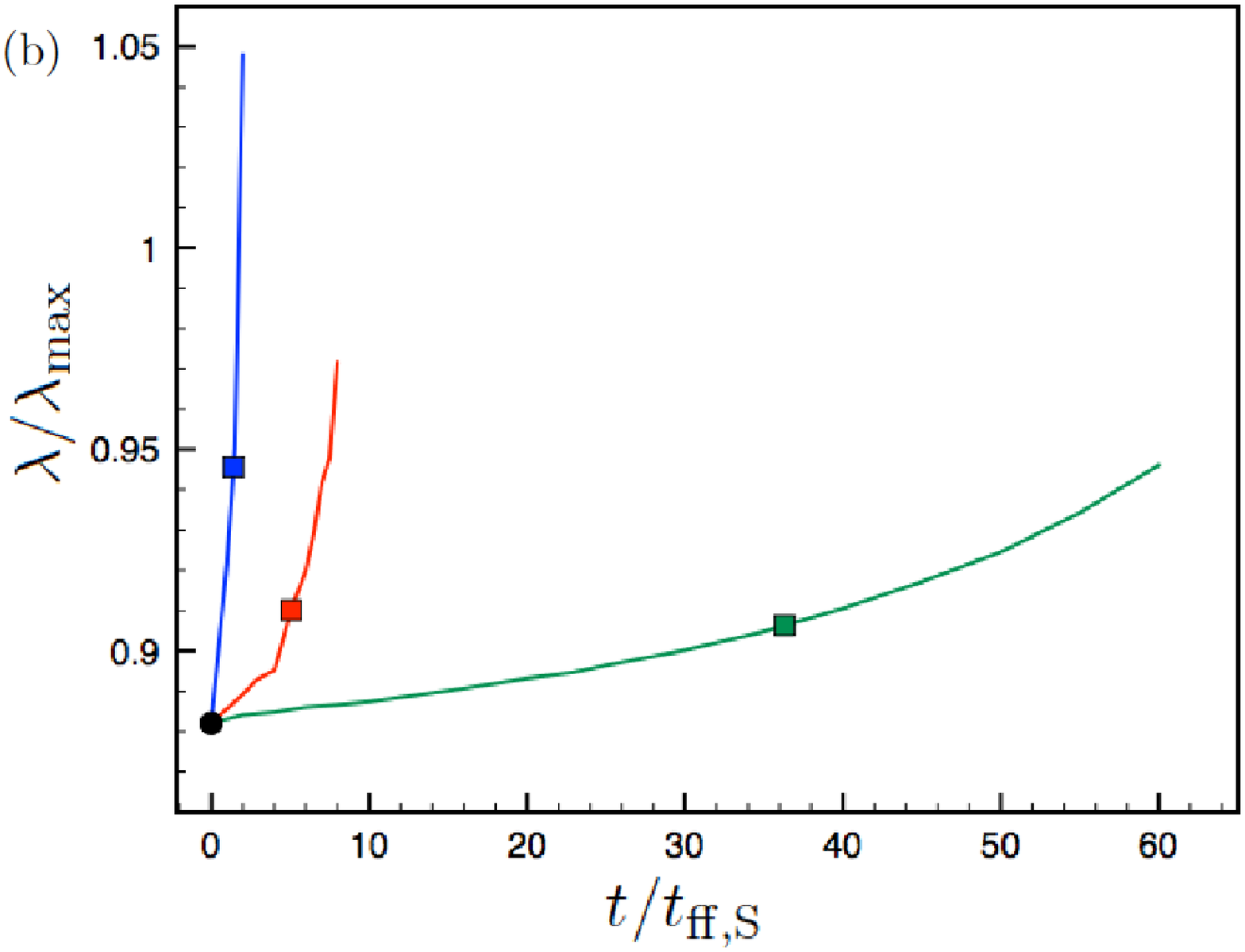}
  \caption{(a) Normalised central density, $\rho/\rho_{\rm S}$ as a function of 
  time for ambipolar diffusion regulated collapse of  model C3b. 
  (b) Ratio of line mass to critical line mass with the black dot the value 
  of the equilibrium profile. The squares show the ratio when the central density
  is 1000 $\rho_{\rm S}$.
  The blue, red
  and green lines show the collapse for $X = 10^{-8}, 10^{-7}$ and $10^{-6}$
  respectively. The black line shows the collapse if no magnetic fields are 
  present.
  }
  \label{fig:C3b evolution}
  \end{figure}

\subsection{Thermally super-critical filaments}
All other models are thermally super-critical and continue to collapse 
gravitationally until all the gas mass is in a few grid cells. 
Figure~\ref{fig:C3b evolution}a shows the evolution of the central density for 
different values of $X$. As $X$ decreases, the timescale for collapse also 
decreases, that is the collapse timescales are approximately $56.6 t_{\rm ff,S}, 
7 t_{\rm ff,S}$ and $1.71 t_{\rm ff,S}$ for $X = 10^{-6}, 10^{-7}$ and $10^{-8}$ 
respectively. 

This dependency can be understood by considering the flux loss timescale, 
$t_{\phi}$, for gravitational collapse. \citet{1991ApJ...371..296M} show that 
the flux loss timescale is given by 
\begin{equation}\label{ambiref}
t_{\phi} = \frac{\nu_{\rm ff}t_{\rm ff}}{1.4} \approx 2.14\frac{\chi K_{\rm ni}}{\pi G},
\end{equation}
where $\nu_{\rm ff} = t_{\rm ff}/t_{\rm ni}$ is the collapse retardation factor, 
$t_{\mathrm{ni}} = 1/\rho_i K_{\rm ni}$ the average collision time between neutrals 
and ions and $\chi$ the ionisation fraction at the initial time. As the fractional
ionisation is given by $\chi = X\rho_n^{-1/2}$, we find that 
the collapse time is proportional to $X$ if the collapse time substantially exceeds the 
free-fall time.  Using the central density of the filament for $\rho_n$, we find 
$t_{\phi} \approx 125 t_{\rm ff,S}$ for $X = 10^{-6}$, $\approx 12.5 t_{\rm ff,S}$ 
for $X = 10^{-7}$ and $\approx 1.25 t_{\rm ff,S}$ for $X = 10^{-8}$. 
These estimated values agree within a factor of two with the numerically derived 
values and thus explain the near-proportional decrease in collapse timescale 
of the filament.
However, it is clear that the proportionality needs to break down at some point 
as the gravitational contraction rate cannot exceed the free-fall collapse rate. 
Numerically the minimum collapse timescale is derived by instantaneously removing the 
magnetic flux support, that is setting the magnetic field values to zero and we find 
$\approx 0.9 t_{\rm ff,S}$ (see black line in Fig.~\ref{fig:C3b evolution}).
From Eq.~\ref{ambiref}, we expect this to happen when $\nu_{\rm ff} \lesssim 1.4$
or $X \lesssim 8\times 10^{-9}$. 
\citet[][]{2012ApJ...761...67B} find a similar dependence on the ionisation fraction
for the collapse time of planar sheets. A linear stability analysis shows 
that the collapse time for sheets with mass-to-flux ratios equivalent to our models 
decreases linearly until the average ion-neutral collision time becomes comparable 
to the free-fall collapse time of the thin sheet.   

\begin{figure}
  \centering
  \includegraphics[width=9cm]{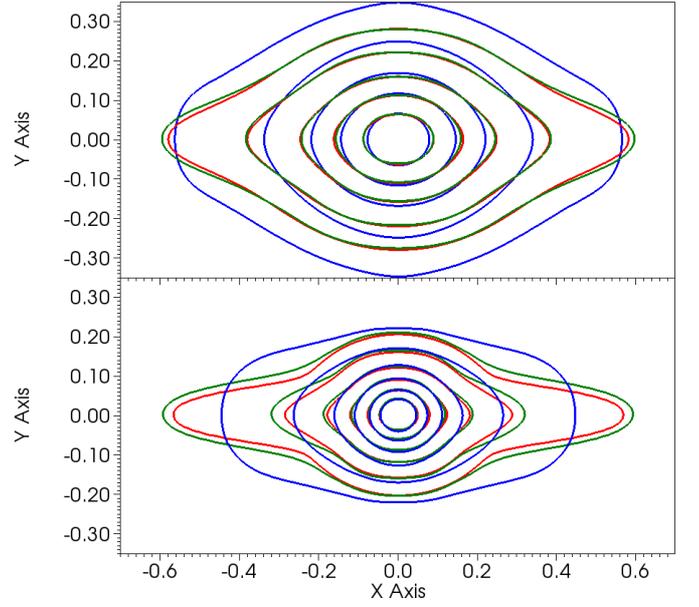}
  \caption{Normalised density contours, $\rho/\rho_{\rm S}$, for $X = 10^{-8}$ (blue), 
  $X = 10^{-7}$ (red) and $X = 10^{-6}$ (green). The top panel 
  shows the density configuration for a central density of $\approx 250 \rho_{\rm S}$, 
  while the bottom shows it for $\approx 1000 \rho_{\rm S}$. The contour line show  
  normalised densities of 1, 3, 10, 30, 100 and 300.}
  \label{fig:C3b rho configurations}
  \end{figure}

\begin{figure}
  \centering
  \includegraphics[width=9cm]{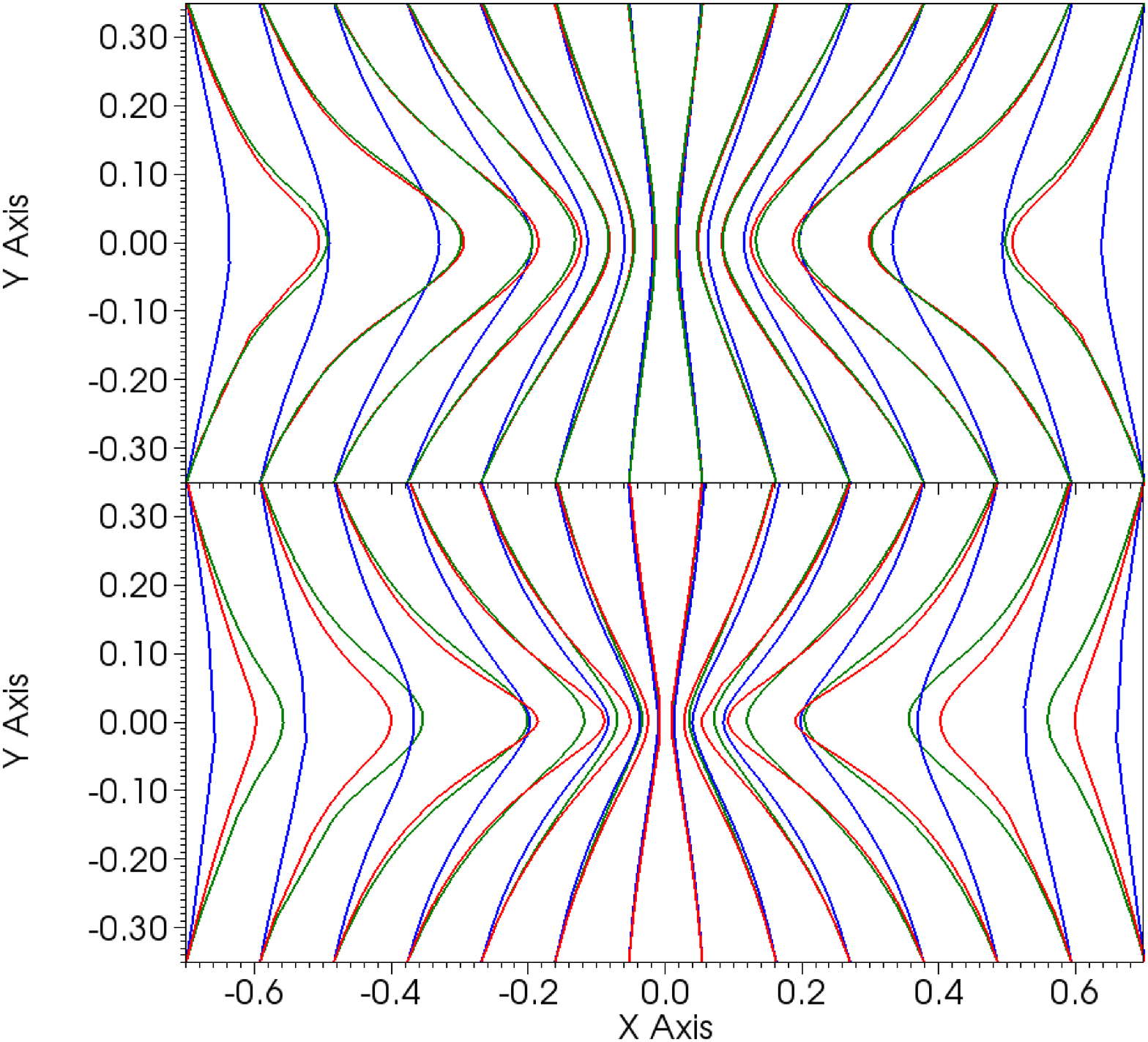}
  \caption{Magnetic field lines for $X = 10^{-8}$ (blue), $X = 10^{-7}$ (red) and 
  $X = 10^{-6}$ (green) when the central density of the filament is 
  $\approx 250 \rho_{\rm S}$ (top panel) and when it is $\approx 1000 \rho_{\rm S}$
  (bottom panel).}
  \label{fig:C3b B configurations}
  \end{figure}

This break is also observed in the evolution in the distribution of gas and the magnetic field 
in the filament. Figures~\ref{fig:C3b rho configurations} and \ref{fig:C3b B configurations} 
show the density profile and magnetic field configuration for $X = 10^{-6}, 10^{-7}$
and $10^{-8}$ when the central density has a value of $250 \rho_S$ and $1000 \rho_S$.
(We note that these densities are attained at different times for each model). These
densities are chosen to represent an instant during the linear ($250 \rho_S$) and
non-linear ($1000 \rho_S$) phase of the filament collapse. For $X = 10^{-6}$ and $10^{-7}$, 
the profiles and field structure evolve identically, albeit at different collapse rates, 
indicating a self-similar evolution. The magnetic field for these values of $X$ maintains 
its hourglass shape throughout the collapse and, during the linear phase, even preserves the 
field from the equilibrium distribution. Then, from Eq.~\ref{eq:Bfield} follows  
\begin{equation}\label{eq:neutral velocity}
{\bf v_n} - \frac{r_a}{B^2} (\nabla \times {\bf B}) \times {\bf B} \approx 0.
\end{equation} 
As the magnetic field structure is independent of $X$ (if $X \geq 10^{-7}$), 
the neutral velocity varies as $r_a$ which is inversely proportional to $X$
(see Eq.~\ref{eq:ra}). This again explains why the collapse rate is 
proportional to $X$. The neutral speeds are small compared to the sound speed, so that 
the evolution is slow and much slower than the free-fall collapse, that is quasi-static. 
As mentioned earlier, for these values of $X$, $\nu_{\rm ff}$ is large. The collision
timescale between neutrals and ions is therefore much shorter than the free-fall 
timescale, so that the neutrals are still strongly coupled to the magnetic field 
and the ambipolar diffusion length scale, that is $\lambda_{\rm AD} = \pi v_{\rm A} t_{\rm ni}$, 
is smaller than the Jeans' length $\lambda_J = a t_{\rm ff}$. 
So, the collapse of the cloud, although driven by gravity, is quasi-static and magnetically 
regulated. We will refer to this as gravitationally-driven magnetically-regulated ambipolar 
diffusion.

The evolution for $X = 10^{-8}$ differs from the self-similar solution, although
the collapse is still initiated by 
gravitationally-driven magnetically-regulated ambipolar diffusion.
The collision timescale between neutrals and ions is not too dissimilar to the free-fall 
timescale. The neutrals are therefore weakly coupled to the magnetic field and the magnetic 
field is able to straighten itself quickly (see Fig.~\ref{fig:C3b B configurations}).
This is particularly apparent in the outer regions of the filament where 
$\lambda_{\rm AD} > \lambda_{\rm ff}$. We note that, in Fig.~\ref{fig:C3b rho configurations}, 
the contour line for $\rho_S = 1$ is distinctly different than for the self-similar collapse. 
As the neutrals easily diffuse across the magnetic field lines, the neutral velocities 
for the collapse are much higher than before and close to the sound speed. 
The cloud collapse is thus a dynamical, gravity dominated process and 
is referred to as gravitationally-driven and -dominated ambipolar diffusion.
(As both modes are gravitationally driven, we will drop the adjective
gravitationally-driven for the remainder of the paper).

Some similarities remain for both evolutionary paths, that is the density distribution
in the central region of the filament is independent of the ionisation coefficient $X$.
For the bottom panel of Fig.~\ref{fig:C3b rho configurations}, the density 
distributions above $\rho > 30 \rho_{\rm S}$ are identical. As the central region of
the filament loses magnetic support, the dynamics is determined by gravity and thermal
pressure gradients alone. Furthermore, the neutral velocities within the central region
are much smaller than the sound speed, so that a near static equilibrium is achieved there.
Within the contour of $300\rho_{\rm S}$, the relative difference between the
density distribution and the Ostriker density profile (Eq.~\ref{ost}) is less than 20\%.
So more than a third of the total line mass lies within a region accurately described
by a hydrodynamical equilibrium. 

Although the filament loses magnetic support, it is important to realise that 
not much magnetic flux is lost from the filament due to ambipolar diffusion. Even 
during the non-linear collapse phase, that is when the central density reaches 
$1000\rho_{\rm S}$, the magnetic flux per unit length has decreased by only 5\% of its  
initial value for $X = 10^{-8}$ and less than 1\% for $X \geq 10^{-7}$
(see Fig.~\ref{fig:C3b evolution}b). 
The magnetic flux is actually redistributed within the filament. As 
magnetic flux is transported from the centre towards the envelope, the central regions 
collapse while the outer layers around the midplane remain in place. This can be seen 
in Fig.~\ref{fig:C3b rho configurations} especially for $X \geq 10^{-7}$.

\begin{figure}
  \centering
  \includegraphics[width=9cm]{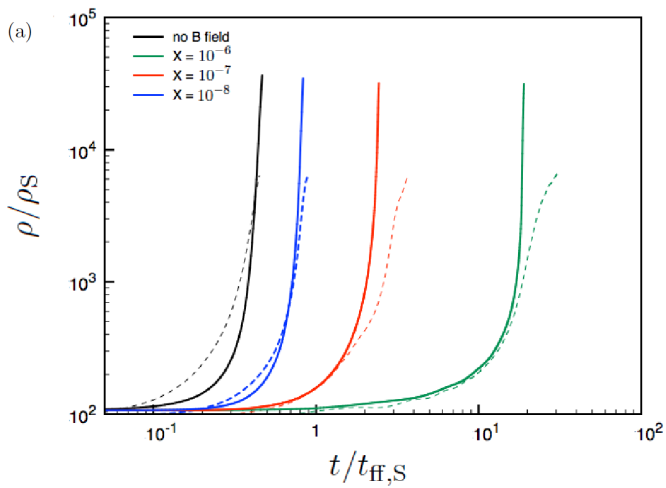}
  \includegraphics[width=8.5cm]{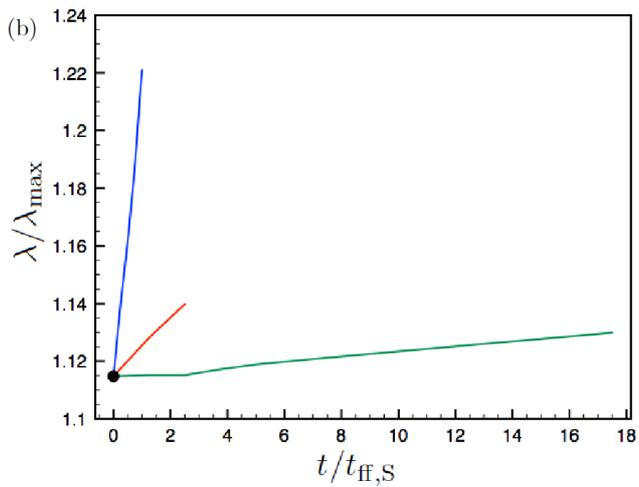}
  \caption{
  Same figures as for Fig.~\ref{fig:C3b evolution} but for model D2.
  The dashed lines show the evolution for model C3b from 
  Fig.~\ref{fig:C3b evolution} with the time renormalised so that the numerical  
  time scale for instantaneous flux loss is the same as for model D2.}
  \label{fig:D2 evolution}
  \end{figure}

\begin{figure}
  \centering
  \includegraphics[width=9cm]{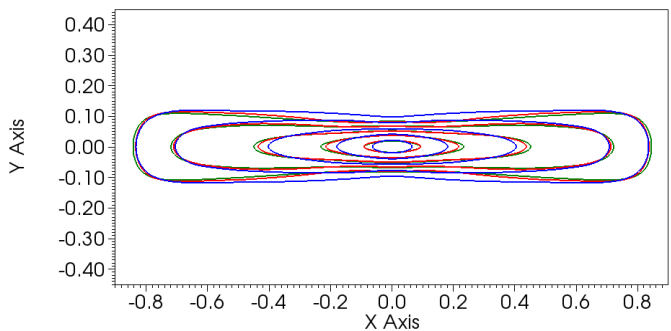}
  \caption{Normalised density contours, $\rho/\rho_{\rm S}$, for $X = 10^{-8}$ (blue), 
  $X = 10^{-7}$ (red) and $X = 10^{-6}$ (green) for model D2 when the central 
  density is $\approx 200 \rho_{\rm S}$.  The contour line show  normalised densities 
  of 1, 3, 10, 30 and 100.}
  \label{fig:D2 rho configurations}
  \end{figure}

While the above describes the collapse of model C3b for different ionisation 
coefficients, the results are generally valid for the other thermally super-critical 
models. Figures~\ref{fig:D2 evolution} shows the evolution of the central density 
and the ratio of the line mass to the critical line mass (Eq.~\ref{eq:tomisaka}) 
for model D2 as the filament is collapsing.  When overlaid with the evolution for model C3b 
renormalised in time so that the numerical time scales for the instantaneous flux loss 
model are the same, the trends are seen to be identical. This is expected
as the collapse retardation factor only depends on the ionisation coefficient $X$ and
not on for example density. A similar argument holds for the density distribution. Again,
for $X \lesssim 10^{-8}$ the distribution differs as the collapse is 
regulated by 
gravitationally-dominated
ambipolar diffusion. For higher ionisation
coefficients, 
magnetically-regulated
ambipolar diffusion forces the filament to 
undergo quasi-static collapse (see Fig.~\ref{fig:D2 rho configurations}).

\section{Decaying turbulence and ambipolar diffusion}\label{sect:turbulent}
For high ionisation coefficients the gravitational collapse regulated by ambipolar 
diffusion is still a slow process compared to free-fall collapse. However,
turbulence accelerates ambipolar diffusion in the absence of self-gravity
\citep[e.g.][]{2004ApJ...603..165H, 2012ApJ...744...73L}. \citet{2002ApJ...570..210F} 
show analytically that there is an increase in the ambipolar diffusion rate, and 
thus the collapse rate, of a factor of a few. 
Three-dimensional simulations of sub-critical thin sheets further corroborate 
this finding as the timescale for core formation shortens when velocity perturbations 
are considered in addition to ambipolar diffusion \citep[][]{2011ApJ...728..123K}.

In order to study the effect of turbulence on the ambipolar diffusion rate, 
we perturbed the equilibrium distribution of model C3b by adding velocity
perturbations $\delta v_{x,y}$ appropriate for turbulence to the neutral velocity. 
We used an approach described in \citet{1999ApJ...524..169M}, although 
we do not subsequently drive the turbulence. The velocity 
perturbations are generated by assigning an amplitude and a phase in Fourier 
space and transforming them back into real space. While the phase is a random 
number between 0 and $2\pi$, the amplitude is drawn from a Gaussian distribution 
around zero and a deviation given by $P(k) \propto k^{-2}$, where 
$k = L_d/\lambda$ is the dimensionless wave number ($L_d = 2$ is the largest driving 
wavelength). We assumed a set of wave numbers ranging from $1 \leq \sqrt{k_x^2 + k_y^2} \leq 100$. 
We also ensured that there is no net momentum input into the filament, that is 
$\int{\rho\delta v_{x,y} dV} = 0$. Then we normalised the amplitude of the velocity 
perturbations so that the turbulent velocity field has a rms sonic Mach number, 
$M_{\rm rms}$, of 10, 3 or 1. These values were chosen to see an effect of the 
turbulent motions within the filament even though the turbulence decays.

  \begin{figure}
  \centering
  \includegraphics[width=9cm]{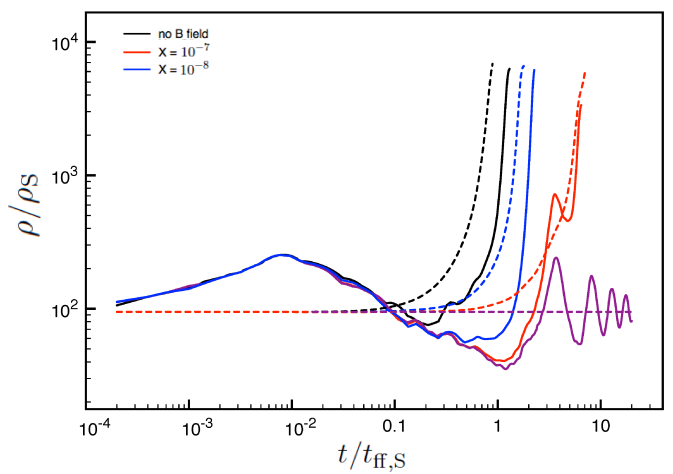}
  \caption{Similar figure as Fig.~\ref{fig:C3b evolution}, but with a turbulent
    velocity field of $M_{\rm rms} = 10$ added to the equilibrium filament. The purple 
    line shows the evolution of the ideal MHD filament. The dashed lines show the 
    quiescent evolution of model C3b. 
  }
  \label{fig:C3b turbulent evolution}
  \end{figure}

Figure~\ref{fig:C3b turbulent evolution} shows the evolution of the maximum density 
in the filament as a function of time for different ionisation coefficients $X$
and for ideal MHD with and without turbulence (The initial rms Mach number is 10).
Initially the turbulent motions compress the gas locally leading to an increase
in the maximum density by a factor of a few ($\approx 2-3$). However, the overall effect
of the turbulence is to oppose self-gravity. Thus, the filament expands lowering the 
maximum density. At the same time the turbulence is decaying and, around $0.1 t_{\rm ff,S}$,
enough turbulent support is removed, that is $M_{\rm rms}$ reduces to $\approx 0.4$, 
for the filament to collapse in the absence of magnetic fields. The turbulence 
significantly slows down the gravitational collapse, that is by a factor 1.6.  

  \begin{figure*}
  \centering
  \includegraphics[width=\textwidth]{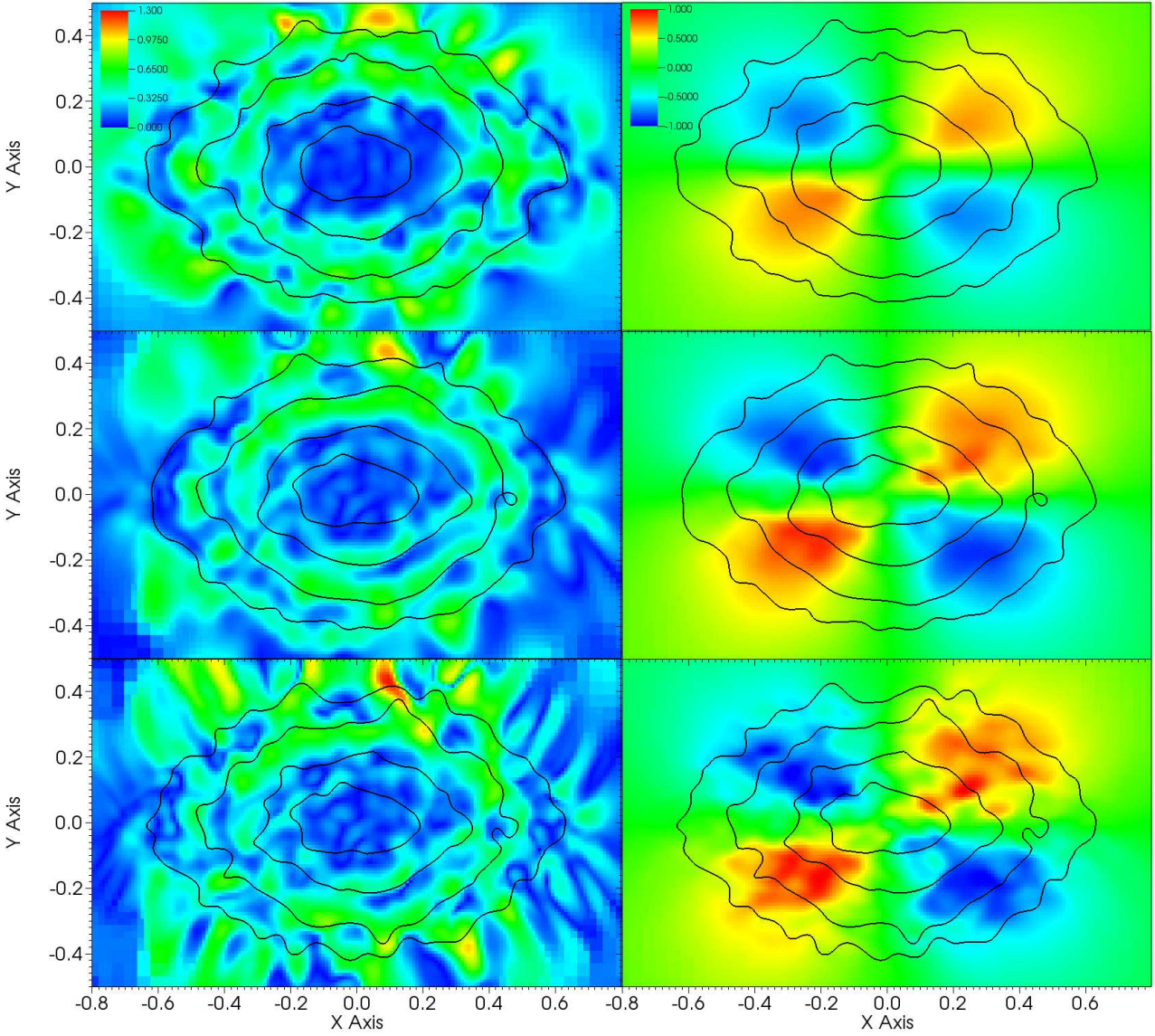}
  \caption{Velocity magnitude (left) and $x$-component of the magnetic field (right) 
    after $0.3 t_{\rm ff,S}$ for $X = 10^{-8}$ (top),
    $X = 10^{-7}$ (middle) and ideal MHD (bottom) with a initial turbulent velocity
    field of $M_{\rm rms} = 10$. The contour lines show 1, 3, 10 and 30 $\rho_{\rm S}$. 
  }
  \label{fig:turb dissipation}
  \end{figure*}

The turbulence is still sufficient to support the filaments if a magnetic field is present. 
For magnetic-field supported filaments both magnetic support and turbulent support need to be 
removed. The ambipolar diffusion helps with both. As in the quiescent case ambipolar 
diffusion redistributes the magnetic flux within the filament so that most of the flux
is in the envelope and not in the centre. The turbulent velocity field generates larger 
magnetic field gradients than in the quiescent case. It thus accelerates the diffusion 
of the magnetic field on the small scales (see right-hand term in Eq.~\ref{eq:Bfield})
and eventually smooths the large scale structure \citep{2004ApJ...603..165H}. At the same 
time, ambipolar diffusion dissipates perturbations with wavelengths below 
$\lambda_{\rm AD}$ \citep{1991ApJ...371..296M,2008A&A...484..275V} so that, 
also, turbulent support is removed.  The dissipation of small-scale velocity 
perturbations is displayed in Fig.~\ref{fig:turb dissipation}. This figure shows the 
magnitude of the velocity field and the $x$-component of the magnetic field for different 
ionisation coefficients at $t = 0.3 t_{\rm ff,s}$. At this time the models are still 
similar in their density profile (e.g. their central density is similar as can be seen
in Fig.~\ref{fig:C3b turbulent evolution}), but show variation in the velocity and 
magnetic field. The difference can also be quantified. While the difference in rms 
velocity is small when considering the entire filament ($M_{\rm rms}$ lies between 
0.262 and 0.285), the rms velocity within the highest density contour (> 30$\rho_{\rm S}$) 
varies significantly between $X = 10^{-8}$, $X = 10^{-7}$ and ideal MHD with values 
of $M_{\rm rms} = 0.087$, 0.129 and 0.148 respectively. A lower rms velocity indicates 
less turbulent support. The presence of the velocity perturbations is also seen in the 
$x$-component of the magnetic field. As the velocity perturbations dissipate more slowly 
for a higher ionisation coefficient and the gas is also more strongly coupled to the 
field, turbulence distorts the magnetic field to a higher degree. 

  \begin{table}
  \caption{Ratio of collapse time for ambipolar-diffusion regulated collapse with 
        different ionisation coefficients to the collapse time if the 
	magnetic field is instantaneously removed with changing levels of 
	turbulence. }
  \label{table:ratios time}      
  \centering                         
  \begin{tabular}{c c c}        
  \hline\hline                 
  $M_{\rm rms}$ & $X = 10^{-8}$&$X = 10^{-7}$\\
  \hline                       
	0 & 1.982 & 8.078\\
	1 & 1.919 & 8.533 \\
	3 & 1.921 & 8.292\\
        10 &  1.768 & 6.071\\
  \hline                                   
  \end{tabular}
  \end{table}

Unfortunately, it is not possible to quantify the acceleration of the ambipolar diffusion 
by the turbulence. The filament's evolution depends strongly on the amount of turbulence 
injected and disentangling the associated effects from the ambipolar diffusion acceleration is not 
feasible. However, we can compare the time scale of the ambipolar-diffusion regulated 
collapse with the collapse time for instantaneously removal of the magnetic field to 
quantify the combined effect (see Tab.~\ref{table:ratios time}). The ratios for 
$M_{\rm rms} = 10$ show that turbulence indeed speeds up the ambipolar diffusion process
as in the simulations of \citet[][]{2011ApJ...728..123K}.
However, note from Fig.~\ref{fig:C3b turbulent evolution} that the actual time for 
collapse increases when turbulence is present. Turbulence adds extra support against 
self-gravity, but it happens both in the models with and without magnetic fields. 
Also, we should take into account that the ratios are only upper values as most of the 
turbulence decays well before the collapse is finished, especially for $X = 10^{-7}$.
To fully understand the interplay of ambipolar diffusion and turbulence, driven-turbulence
models are needed.

\section{Discussion and conclusions} \label{summary}
In this paper we have investigated the effect of ambipolar diffusion and decaying 
turbulence on infinitely long, isothermal, magnetically sub-critical filaments in two dimensions. 
Magnetohydrostatic equilibrium filaments in pressure-equilibrium with the external medium 
are generated numerically in ideal MHD as initial conditions. These equilibria reproduce the 
analytic profiles of \citet{2014ApJ...785...24T} and, by perturbing the equilibria with decaying 
velocity perturbations, we find that these equilibrium filaments are dynamically stable.

By using a multifluid AMR MHD code, we then follow the response of the equilibrium filament 
to ambipolar diffusion. Due to the gradients of the magnetic field, ambipolar diffusion 
initiates the filament's contraction. For thermally sub-critical filaments this contraction is 
halted when a new equilibrium is reached. Magnetic support is lost, with flux loss rates 
increasing inversely proportional to the ionisation coefficient $X$, but thermal pressure gradients 
are enough to balance gravitational forces. The new equilibrium is the hydrostatic profile 
described by \citet{1964ApJ...140.1056O}. 

For thermally super-critical filaments the filament contains enough mass to overcome 
thermal pressure forces and to collapse gravitationally. The collapse rate depends on the 
flux loss rate and is, as for the sub-critical filaments, thus inversely proportional to the 
ionisation coefficient $X$. It is important to realise that ambipolar-diffusion regulated
collapse solely depends on $X$ and no other variable such as for example density, magnetic field strength 
or external pressure. Two models with completely different properties, that is C3b and D2, show 
the same collapse times for the various values of $X$ when normalised to the collapse time for 
instantaneous magnetic flux removal. 

Two gravitationally-driven ambipolar diffusion regimes are observed:
a magnetically-regulated one for $X \geq 10^{-7}$ and a gravitationally-dominated one for $X \lesssim 10^{-8}$
in agreement with \citet{1991ApJ...371..296M}. The former arises because the collision time between 
neutrals and ions is much shorter than the free-fall time (or $\lambda_{\rm AD} < \lambda_{\rm ff}$). 
Then the neutrals are strongly coupled to the magnetic field and the ambipolar diffusion is 
regulated by the magnetic field.
The collapse is quasi-static with neutral velocities much smaller 
than the sound speed. In the latter regime, the ion-neutral collision time becomes comparable or 
longer than the free-fall time and $\lambda_{\rm AD} \geq \lambda_{\rm ff}$. The neutrals
are then weakly coupled and high neutral velocites are attained. 

Ambipolar-diffusion regulated collapse is a slow process compared to free-fall collapse, especially
in the 
magnetically-regulated
regime. Numerical simulations of non-gravitating clouds show that turbulence
enhances ambipolar diffusion by a factor of a few \citep{2004ApJ...603..165H, 2012ApJ...744...73L}. 
When the equilibrium filament is perturbed by adding a decaying turbulent velocity field, 
we find that the ambipolar-diffusion collapse times decrease when compared to the collapse
time for instantaneous magnetic flux loss. The actual time scales increase as 
the turbulent motions provide additional support to the filament. The effect of the turbulence
on the ambipolar diffusion is to speed up the diffusion rate as larger magnetic field gradients 
are generated, while ambipolar diffusion dissipates the turbulence below the ambipolar diffusion
length scale. Because we only study the effect of decaying turbulence, we cannot disentangle the 
combined effect of turbulence and ambipolar diffusion, but the largest effect is observed for 
the lowest ionisation coefficient and the highest turbulent intensity. 

Other effects potentially enhance the collapse rate of the magnetised filament further. 
In addition to ambipolar diffusion, turbulent magnetic reconnection can be an efficient 
diffusion process for the magnetic field \citep[e.g.][]{1999ApJ...517..700L,2010ApJ...714..442S}. 
Also, our models are restricted to 2D. \citet{1991ApJ...373..169M} has shown that geometry,
that is the dimensionality of the problem, plays an important role in the fragmentation process
due to ambipolar diffusion. In a subsequent paper we will extend these filaments to three 
dimensions.

\section*{Acknowledgments}
We thank the anonymous referee for his/her insightful comments that have 
improved the paper.
The calculations for this paper were performed on the DiRAC Facility jointly funded by STFC, 
the Large Facilities Capital Fund of BIS and the University of Leeds. CAB, SAEGF and TWH 
are supported by a STFC consolidated grant. The data presented in this paper is available 
at http://doi.org/10.5518/72.

{}

\end{document}